# Spin-lattice relaxation phenomena in the magnetic state of a suggested Weyl semimetal CeAlGe

Karan Singh and K. Mukherjee

School of Basic Sciences, Indian Institute of Technology Mandi, Mandi 175005, Himachal Pradesh, India

## ABSTRACT

In this work we report the results of DC susceptibility, AC susceptibility and related technique, resistivity, transverse and longitudinal magnetoresistance and heat capacity on polycrystalline magnetic semimetal CeAlGe. This compound undergoes antiferromagnetic type ordering around 5.2 K ($T^1$). Under application of external magnetic fields, parallel alignment of magnetic moments is favored, above 0.5 Tesla. At low field and temperature, frequency and AC field amplitude response of AC susceptibility indicate to the presence of spin-lattice relaxation phenomena. The observation of spin-lattice interaction suggests to the presence of Rashba-Dresselhaus spin-orbit interaction which is associated with inversion and time reversal symmetry breaking. Additionally, presence of negative and asymmetric longitudinal magnetoresistance indicates anomalous velocity contribution to the magnetoresistance due to Rashba-Dresselhaus spin-orbit interaction which is further studied by heat capacity.

1. **Introduction**

Non-magnetic semimetals have attracted significant attention, both, theoretically and experimentally among physicists working in the area of topological materials. In a semimetal, conduction and valence bands cross each other in the Brillouin zone. This crossing persists under various perturbations and preserves certain crystalline symmetry. Such systems exhibit novel topological properties like quantum anomalies [1-4]. These anomalies are characterized by violation of classical symmetry due to quantum effect in the presence of external field [5-7]. In this context, the breaking of either inversion or time reversal symmetry forms the Weyl points at the Fermi surface, and this phenomenon has attracted significant attention in semimetals. In such cases, Fermi surface consists of open arcs connecting the projection of bulk Weyl points onto the surface Brillouin zone, instead of closed loops. It has remarkable nontrivial topological properties like integer momentum space invariance [8-9]. This invariance arises from the singularities in the electronic structure where the different bands touch, as isolated points [10]. The symmetry breaking occurs either by natural or external perturbation in form of spin-orbit coupling thereby, leading to interesting topological gapped or gapless states [11-13]. In recent years, magnetic semimetal has emerged as an exclusive topic of research due to interplay of magnetism and topology [14-16]. In this prospective, a variety of compounds like CeSb, $Ce_3Bi_4Pd_3$, CeSbTe, $Co_3Sn_2S_2$ and $Co_2MnGa$ which show the existence of Weyl/Dirac nodes in the magnetic ordered phase has been investigated [16-20]. The presence of localized magnetic moment shifts the Weyl/Dirac nodes leading to an unusual response in the magnetization [19, 21]. Generally, Weyl nodes are noted in a non-centrosymmetric crystal structure. In centrosymmetric crystal structure, breaking of product symmetry (inversion and time reversal) can also allow the Weyl nodes [22]. It is reported that, inversion symmetry maps the Weyl nodes with opposite chirality whereas the time reversal symmetry maps the Weyl nodes without changing its chirality [22]. In the presence of spin-orbit coupling, Zeeman energy split the band in spin up and spins down state, which is further mixed by the Rashba-Dresselhaus effect [23]. The Rashba spin-orbit interaction arises from the structural inversion asymmetry whereas the Dresselhaus spin-orbit interaction arises from the bulk inversion asymmetry [23]. Rashba-Dresselhaus spin-orbit interaction results in the magnetic texture of band electrons. This phenomenon is known as spin momentum locking and it produces a spin torque which may induce a spin relaxation [24-25].

In this context, the compound CeAlGe is interesting. It has been reported that this compound orders ferromagnetically [26] and theoretical studies have predicted a Weyl semimetal state due to breaking inversion and time reversal symmetry [27]. However, recent experimental results of a single crystal CeAlGe report the magnetic semimetal behavior along with antiferromagnetic ordering [28, 29] which contradicts the results of Ref [26] and [27]. In Ref [30], it has been reported that in this compound there is a presence of some parallel alignment of magnetic moments in an antiferromagnetic spin matrix. Therefore, it would be interesting to identify the intrinsic magnetic state and also to see whether there is some evidence of Weyl state in this polycrystalline compound.

Hence, with this aim, in this manuscript we report the results of DC susceptibility, AC susceptibility and related technique, resistivity, transverse and longitudinal magnetoresistance and heat capacity on the polycrystalline magnetic semimetal CeAlGe. This compound undergoes an antiferromagnetic type ordering around 5.2 K at low applied magnetic fields. As the magnitude of applied field is increased, parallel alignment of magnetic moments is favored above 0.5 Tesla. A detailed investigation through AC susceptibility reveals the presence of spin-lattice relaxation in this compound. The observed spin-lattice relaxation is a signature of the presence of Rashba-Dresselhaus spin-orbit interaction. Further, the existence of negative and asymmetric longitudinal magnetoresistance provides the evidence of the anomalous velocity contribution to the magnetoresistance due to Rashba-Dresselhaus spin-orbit interaction.

2. **Experimental details**

The compound CeAlGe is prepared by arc melting stoichiometric amounts of respective high purity (>99.9%) elements in an atmosphere of argon. For better homogeneity, the ingot is re-melted a number of times by turning over each time. The weight loss after the final melting is less than 1%. After melting, the ingot is wrapped in tantalum foil and sealed in an evacuated quartz tube. The compound is annealed at 650 $^0$C for 48 hours and quenched in ice water. The compound thus obtained is characterized by x-ray diffraction (XRD) using Rigaku smart lab diffractometer using monochromatized Cu K$\alpha_1$ radiation at room temperature. Figure 1 shows the Rietveld refined (using Fullprof suite) powder x-ray diffraction pattern of this compound at room temperature. The compound crystallizes in tetragonal structure and is in single phase. The structure can be refined by two space group either: *I*4$_1$/a*md* (centrosymmetric) or *I*4$_1$*md* (non-centrosymmetric). We tried to analyze the pattern using both space groups. It is observed that *R*-factor and goodness of fit are smaller for *I*4$_1$*md* space group which points to the presence of non-centrosymmetric structure. In order to get an idea about the stoichiometry, energy-dispersive x-ray spectroscopy measurement is carried out. The average atomic stoichiometry of the compound is in accordance to the expected values. From Scanning electron microscope images it is revealed that the average crystallite size is of the order of micrometer. Temperature (*T*) and magnetic field (*H*) dependent magnetization (*M*) studies are performed using Magnetic Property Measurement System (MPMS), while temperature and magnetic field dependent heat capacity and resistivity are performed using Physical Property Measurement System (PPMS), both Quantum design, USA. Resistivity, transverse magnetoresistance (magnetic field is perpendicular to the current direction) and longitudinal magnetoresistance (magnetic field is parallel to the current direction) measurements are carried out on pellets of specific shapes. In these measurements, current inhomogeneity is minimized by the procedure mentioned in the supplement information of Ref. [31].

## 3. Result and discussion

**DC Susceptibility study**

Figure 2 (a) shows the temperature dependent DC magnetization divided by magnetic field (*M*/*H*) under zero field cooling (ZFC) and field cooling (FC) conditions in the field range 0.01-0.3 Tesla. It is observed that *M*/*H* increases rapidly and shows a peak around 5.2 K ($T^1$), followed by a hump around 3.0 K (possibly due to some spins rearrangement) at 0.01 Tesla. The bifurcation between ZFC and FC curves start from $T^1$. As the magnitude of applied field increases the bifurcation between the curves reduces and is suppressed around 0.3 Tesla. Above 0.3 Tesla, *M*/*H* curves saturate below the 5.2 K (shown in figure 2 (b)). Inverse magnetic susceptibility of the compound is fitted with Curie Weiss law in temperature range 50 - 300 K at 0.1 Tesla (inset of figure 2 (a)). Below this temperature a non-linear deviation is observed, which could be due to crystalline electric field effect [32]. The obtained effective moment ($\mu_{eff}$) and Curie Weiss temperature ($\theta_p$) are 2.8 $\mu_B$ and –32 K respectively. The negative value of $\theta_p$ indicates to the dominance of antiferromagnetic interactions. It is also noted that $T^1$ decreases on increasing the magnitude of applied field. Hence, peak around $T^1$ arises due to the antiferromagnetic ordering which is in accordance to ref [28, 29]. Above 0.5 Tesla, there is a presence of some partial parallel alignments of magnetic moments [28, 30]. On further increasing magnetic field, this alignment of magnetic moments increases which is responsible for the increase of saturation temperature of *M/H* curve to higher temperature (as shown in the inset of figure 2(b)). In order to shed more light, the magnetic field response of isothermal magnetization *M* at different temperatures is measured (shown in Figure 2 (c)). The curve shows the weak hysteresis at temperatures below $T^I$ (shown inset of figure 2 (c)). However, at high field the magnetization tends to saturate due to parallel alignment among the magnetic moments. From above studies, it can be concluded that at low field, below $T^I$, there is some parallel alignment of magnetic moments in an antiferromagnetic spin matrix. This give rise to the Zeeman energy which increases as the magnetic field is increased because high field favors the parallel alignment among the magnetic moments.

**AC susceptibility: Spin-lattice relaxation phenomena**

To further explore the complex magnetic state present in this compound, AC susceptibility study is done. Figure 3 (a) and 3 (b) shows the temperature response real part ($\chi'_{ac}$) and imaginary part ($\chi''_{ac}$) of AC susceptibility measured at 3 ($10^{-4}$) Tesla AC field along with superimposed DC fields of different magnitudes. At 0 Tesla DC field, a sharp peak is noted around $T^I$. With increasing DC field, $T^I$ shifts to lower temperature. Above 0.5 Tesla, the signal of AC susceptibility is suppressed. The above observation is due to the fact that the exchange energy accountable for the development of the antiferromagnetic state is suppressed, above 0.5 Tesla. The signature of the peak is also noted in the $\chi''_{ac}$ due to dissipation of magnetic energy below 0.5 Tesla. Upper and lower inset of the figure 3 (b) show the temperature response of $\chi'_{ac}$ and $\chi''_{ac}$ measured in 3 ($10^{-4}$) Tesla AC field and 0 Tesla DC field under different frequencies. It is observed that with increasing frequency, $T^I$ remains unchanged. This observation rules out the presence of spin freezing mechanism in this compound [33]. Figure 3 (c) shows the DC magnetic field dependent $\chi'_{ac}$ at 7 Hz and AC field of 3 ($10^{-4}$) Tesla. In paramagnetic region (above $T^I$), $\chi'_{ac}$ decreases with increasing magnetic field, in both directions. Below $T^I$ (at 2.5 K), $\chi'_{ac}$ decreases but a peak is observed near 0.39 Tesla, in both directions of fields. These peaks shift to the higher fields on decreasing temperature (shown inset of figure 3 (c)). This observed peak gives an indication of presence of spin dynamic under application of external magnetic field. Hence, our observation indicates toward the existence of an unusual magnetic state in low field and temperature region in this compound.

In this section, we study the frequency response $\chi'_{ac}$ and $\chi''_{ac}$ at fixed temperature, DC and AC field. Figure 4 (a) and inset of 4 (a) shows the frequency dependent normalized (with respect to the value at 500 Hz) $\chi'_{ac}$ and $\chi''_{ac}$ at 7 Hz, 0 Tesla DC and 3 ($10^{-4}$) AC field at different temperature. It is noted that with decreasing frequency (at 1.8 K), $\chi'_{ac}$ increases and a maximum and minimum occur around 125 Hz and 55 Hz, respectively. Below 55 Hz, $\chi'_{ac}$ increases. Similar feature is also noted at other temperatures. For $\chi''_{ac}$, a minimum and a broad maximum are observed around 76 Hz and 4 Hz, respectively and similar feature is observed for other measurement temperatures. This behaviour of AC susceptibility can be understood on the basis of modified Cole-Cole formalism [21].

$$\chi_{ac}(\omega) = \chi_{ac}(\infty) + [\chi_{ac}(0) - \chi_{ac}(\infty)]/1+(i\omega\tau_0)^{1-\alpha} \quad \ldots (1)$$

where $\chi_{ac}(0)$ is the isothermal susceptibility extrapolated to zero frequency where spin-lattice relaxation is active and $\chi_{ac}(\infty)$ is the adiabatic susceptibility extrapolated to infinite frequency where spin-spin relaxation is active; $\omega = 2\pi f$ is the angular frequency; $\tau_0 = 1/2\pi f_0$ is the characteristic relaxation time and $\alpha$ is the width of frequencies distribution. For an infinite broad distribution, $\alpha$ is 1 and it is 0 for a single relaxation process. The equation (1) can be decomposed in term of real and imaginary part as

$$\chi'_{ac}(\omega) = \chi_{ac}(\infty) + [A_0\{1+(\omega\tau_0)^{1-\alpha}\sin(\pi\alpha/2)\}/\{1+2(\omega\tau_0)^{1-\alpha}\sin(\pi\alpha/2)+(\omega\tau_0)^{2(1-\alpha)}\}] \ldots (2)$$

$$\chi''_{ac}(\omega) = A_0\{(\omega\tau_0)^{1-\alpha}\cos(\pi\alpha/2)\}/\{1+2(\omega\tau_0)^{1-\alpha}\sin(\pi\alpha/2)+(\omega\tau_0)^{2(1-\alpha)}\} \ldots (3)$$

where $A_0 = \chi_{ac}(0) - \chi_{ac}(\infty)$ and positive $A_0$ indicate to the dominance of spin-lattice relaxation behavior [21]. We tried to fit the curves in the figure 4 (a) and inset of 4 (a) with equation (2) and (3), respectively. Due to weak frequency dependence, good fitting of the $\chi''_{ac}(\omega)$ curve is not obtained with equation (3). However, a worthy fitting is obtained for $\chi'_{ac}(\omega)$ in low frequency region (below 76 Hz) for 1.8, 3.6 and 5.4 K (shown in figure 4 (a)). A positive $A_0$ is obtained implying that $\chi_{ac}(0)$ is greater than $\chi_{ac}(\infty)$. Hence, it can be said that spin-lattice interaction is dominant below 76 Hz in this compound. The parameter $\alpha$ and $\tau_0$ roughly obtained from fitting are summarized in table 1. These parameters indicate to broad relaxation process in spin-lattice interaction. In high frequency region (above 76 Hz), the change of slope of AC susceptibility curve indicate a competition between spin-lattice relaxation and antiferromagnetic interactions. Here, we would like to mention that the effect of superimposed DC fields (above 0.5 Tesla) on these curves is negligible and the features in AC susceptibility are suppressed.

Furthermore, AC field ($h_{ac}$) dependent studies of $\chi'_{ac}$ and $\chi''_{ac}$ at 7 Hz, 0 Tesla DC field is carried out at different temperatures. Figure 4 (b) and inset of 4 (b) shows the normalized (with respect to the value at 9 ($10^{-4}$) Tesla AC field) $\chi'_{ac}$ and $\chi''_{ac}$ curves at different temperatures. As noted from the figure, above the ordering temperature, $\chi'_{ac}$ varies insignificantly with $h_{ac}$. Below ordering temperature, $\chi'_{ac}$ increases with increasing $h_{ac}$. A change of slope is noted around 5.2 K and a broad maxima is noted around 5.4 K which is centered around 4 ($10^{-4}$) Tesla. $\chi''_{ac}$ also increases with increasing $h_{ac}$ for all temperatures, except in the temperature range 5.2-5.6 K. In this temperature region, a broad maximum is noted. These observations may be arising due to presence of spin-orbit coupling. It is well-known that orbital magnetic moment diminishes the magnetization while spin magnetic moment favor it in presence of external magnetic field. Equal contribution of both of these moments might be responsible for this observed maximum near the magnetic phase boundary. This observation suggest that the magnetic moments interact with lattice via the spin-orbit coupling as it is the key phenomenon responsible for coupling of spin to the lattice [34].

The presence of Zeeman energy splits the bands into spin up and spin down states in the presence of spin-orbit coupling. As a result these states are further interacted by the Rashba–Dresselhaus effect and leads to the breaking of time reversal symmetry [27]. Rashba effect arises due to structural inversion symmetry breaking and split the spin sub-bands, while, Dresselhaus effect arises due to an additional symmetry breaking, resulting in strain induced Dresselhaus spin-orbit interaction. Thus, both Rashba and Dresselhaus effect add an additional interaction in spin-orbit coupling. An inequality between Rashba and Dresselhaus spin-orbit interaction leads to a non-equilibrium condition in the magnetic state [25]. As a result exchange energy associated with the conduction electrons behaving like a torque [35]. This torque acting on magnetic moments results in a spin-lattice relaxation behavior in the magnetic state of this compound.

**Resistivity, transverse and longitudinal magnetoresistance: Evidence of Rashba and Dresselhaus spin-orbit interactions**

Figure 5 (a) shows the temperature dependent resistivity ($\rho$) at 0 Tesla upto 300 K. With decreasing temperature resistivity decreases and an anomaly is noted around $T^I$ (upper inset figure 5 (a)). Lower insert figure 5 (a) shows the temperature dependent resistivity under different applied fields, upto 15 K. It is observed that anomaly near $T^I$ shifts to the right in temperature above 0.5 Tesla, which is analogy with the DC susceptibility results. Figure 5 (b) displays the transverse magnetoresistance (MR) [= {ρ (H) – ρ (0)}/ρ (0)] at different temperatures. From the figure it is noted that MR increase with the increasing magnetic field in both direction at 20 K and 15 K. The positive MR is likely to be arising from the orbital contribution [36]. Below the 15 K, MR decreases with the increasing magnetic field and it is noted that slope of the curve changes below $T^I$. Figure 5 (c) shows the longitudinal MR measured at different temperatures. Similar to transverse MR, longitudinal MR is positive above 15 K and negative below it. It is also noted that the slope changes of MR is insignificant below $T^I$. However, interestingly it is observed that the longitudinal MR is asymmetric (asymmetric behavior is insignificant in transverse MR). The asymmetric MR is calculated using the formula [37-38]:

$$\text{Asym MR} = [\text{MR}(H+) – \text{MR}(H–)]/2 \quad \ldots.. (4)$$

where MR (H+) and MR (H–) are the magnetoresistance for the positive and negative magnetic field, respectively. Inset of figure 5 (c) shows the longitudinal asym MR at selected temperature. It is observed that asym MR increases with increasing magnetic field. The observed asymmetry in MR may arise due to current jetting effect. The current jetting effect arises due to the presence of inhomogeneous conductivity. This can cause a distortion in the current path due to rise in perpendicular component, which flows normal to the major current direction. This component decreases the potential drop between voltage electrodes resulting in the negative and possible asymmetric MR [37]. However, presence of positive MR and insignificant asymmetry in the transverse MR suggest that current jetting effect is minimal. In fact, this mechanism is understood on the basis of anomalous velocity term associated with non-zero Berry curvature [39-40]. Generally, electron velocity, $v$, can be expressed as [37, 41]:

$$\hbar \vec{v} = \nabla_k \varepsilon(k) + e\vec{E} \times \vec{\Omega}(\vec{H},\vec{k}) \quad \ldots.. (5)$$

where $\varepsilon(k)$ is electron energy, $\vec{E}$ is electric field, $\vec{k}$ is wave vector of electron and $\vec{\Omega}(\vec{H},\vec{k})$ is the Berry curvature. The total current $\vec{J}$ can be expressed as in a diffusive transport [37-38]

$$\vec{J} = \int \frac{d^3\vec{k}}{(2\pi)^3}[\vec{v} + e\vec{E} \times \vec{\Omega}(\vec{H},\vec{k}) + \frac{e}{c}(\vec{\Omega}(\vec{H},\vec{k}) \cdot \vec{v})\vec{H}]n_{\vec{k}} \quad \ldots.. (6)$$

where $n_{\vec{k}} = f_{\vec{k}}^0 + g_{\vec{k}}(\vec{H}, \vec{E})$ is electron distribution function, $f_{\vec{k}}^0$ and $g_{\vec{k}}(\vec{H}, \vec{E})$ are equilibrium and non-equilibrium distribution function respectively. In presence of magnetic field ($\vec{H}$), an induced term $g_{\vec{k}}(\vec{H}, \vec{E})$ is added which result in non-vanishing anomalous velocity contribution to the total current. It can said that the presence of Rashba–Dresselhaus spin-orbit interaction (which is associated with the non-zero Berry curvature) is responsible for the non-trivial dependence of $g_{\vec{k}}(\vec{H}, \vec{E})$, which depends on the orientation of $\vec{H}$ and $\vec{E}$ resulting in the anomalous velocity contribution to the MR [42-45]. The above statement can be understood on the basis that with an increase in magnetic field, parallel alignment among magnetic moments increases and it leads to the enhancement of Zeeman energy. This energy associated with the Rashba–Dresselhaus spin-orbit interaction leading to an asymmetry in longitudinal MR, which gives a signature of the time reversal symmetry breaking. As stated before, asymmetry in MR is observed only in current direction parallel to magnetic field, however, considerable change of slope below $T^I$ is noted only for transverse MR. Hence, it can be said that magnetic ordering strength is stronger in the transverse direction and it might be dominate over the anomalous velocity contribution to the MR leading to the observation of insignificant asymmetric MR when the direction of current is perpendicular to magnetic field.

**Heat capacity study**

Figure 6 (a) shows the temperature dependent heat capacity divided by temperature ($C/T$) at different fields. It is noted that a transition appears near $T^I$. With increasing magnetic field, $T^I$ is shifted to the lower temperature. Above 0.5 Tesla, with an increase of magnetic field, the peak broadens and $T^I$ shifts toward higher temperature. In order to extract the electronic ($\gamma$) and phonon ($\beta$) contribution, the following equation is fitted to the 0 Tesla curve above $T^I$ (upper inset of figure 6 (a)):

$$C/T = \gamma + \beta T^2 \quad \ldots (7)$$

The obtained parameters $\gamma$ and $\beta$ are 22 mJ mol$^{-1}$ K$^{-2}$ and 0.26 mJ mol$^{-1}$ K$^{-4}$. The value of $\gamma$ indicates towards the insignificant hybridization of localized magnetic moment with the conduction electrons [46, 47]. Above 0.5 Tesla, $C$ varies as $T^3$ in the low temperature region (where the phonon contribution is negligible). The curves above 0.5 Tesla is fitted with the following equation [17, 48]

$$C = \gamma T + (k_B T/\hbar v^*)^3 k_B \quad \ldots\ldots\ldots (8)$$

where $k_B$ is Boltzmann constant, $v^*$ is the effective Fermi velocity. This relation is proposed for a linear dispersion $\varepsilon(k) = \hbar v^* k$. Due to presence of very small Debye temperature (estimated from $\beta$) it is expected that $\varepsilon(k)$ cannot be due to acoustic phonons. It can be understood on the basis of Rashba-Dresselhaus spin-orbit interaction [24]. Figure 6 (b) shows $T^2$ dependent $C/T$ at selected fields and linear fitting of equation (8) gives the value γ and $v^*$. It is observed that γ decreases while $v^*$ increases with increasing magnetic field (inset of figure 6 (b)) and this feature is similar to that reported in Ref [48]. Hence, it can be said that Rashba- Dresselhaus spin-orbit effect and linear dispersion suggests the presence of symmetry (inversion and time reversal) breaking. As a result, this might be responsible for the spin-lattice relaxation phenomena in magnetic state and such behavior is expected in the vicinity of Weyl nodes [49]. In our case, chiral anomaly is not seen clearly, as it may be hidden due to the interplay of magnetic interaction in this compound. For an unambiguous identification of this hidden signature, ARPES, NMR, x-ray scattering, field dependent neutron diffraction studies, etc. are necessary.

## 4. Conclusion

In summary, our study on CeAlGe reveals that, above 0.5 Tesla and below $T^I$, antiferromagnetic ordering is suppressed and parallel alignment of magnetic moments is favored, while at low field and temperature, spin-lattice relaxation phenomena is noted. The spin-lattice interacts via the Rashba-Dresselhaus spin-orbit interaction due to breaking of symmetry and it results in anomalous velocity contribution to the MR.

**Acknowledgements**

The authors acknowledge IIT Mandi for financial support and the experimental facilities.

**Table 1:** Parameters ($\alpha$ and $\tau_0$) obtained from the AC susceptibility curves fitting with equation (2) in the low frequency region at 0 DC field.

| $T$ (K) | $\alpha$ | $\tau_0$ |
|---|---|---|
| 1.8 K | 0.380±0.018 | 0.027±0.017 |
| 3.6 K | 0.390±0.022 | 0.040±0.026 |
| 5.4 K | 0.398±0.015 | 0.018±0.005 |

**Figures:**

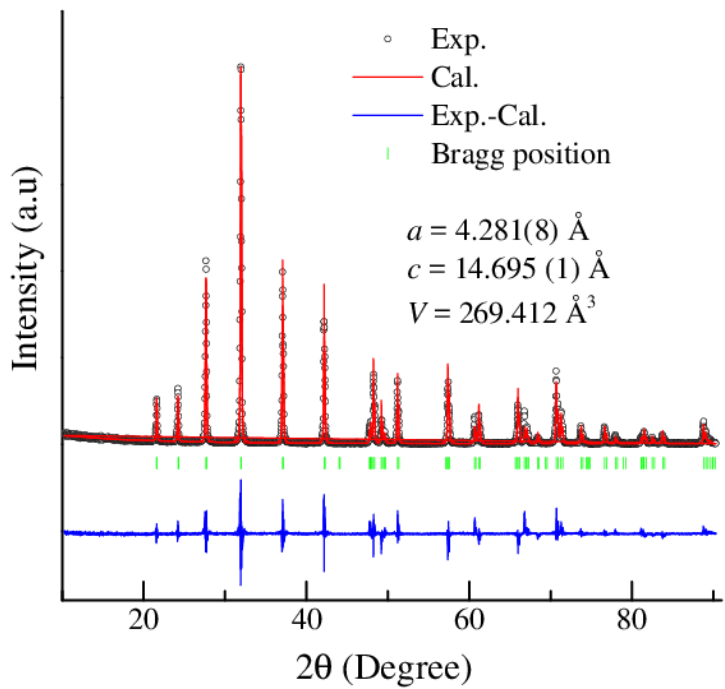

**Figure 1:** Rietveld refined powder XRD pattern of CeAlGe. The difference between the observed and experimental pattern and the Bragg position are also shown. Lattice parameters obtained from the fitting is also represented.

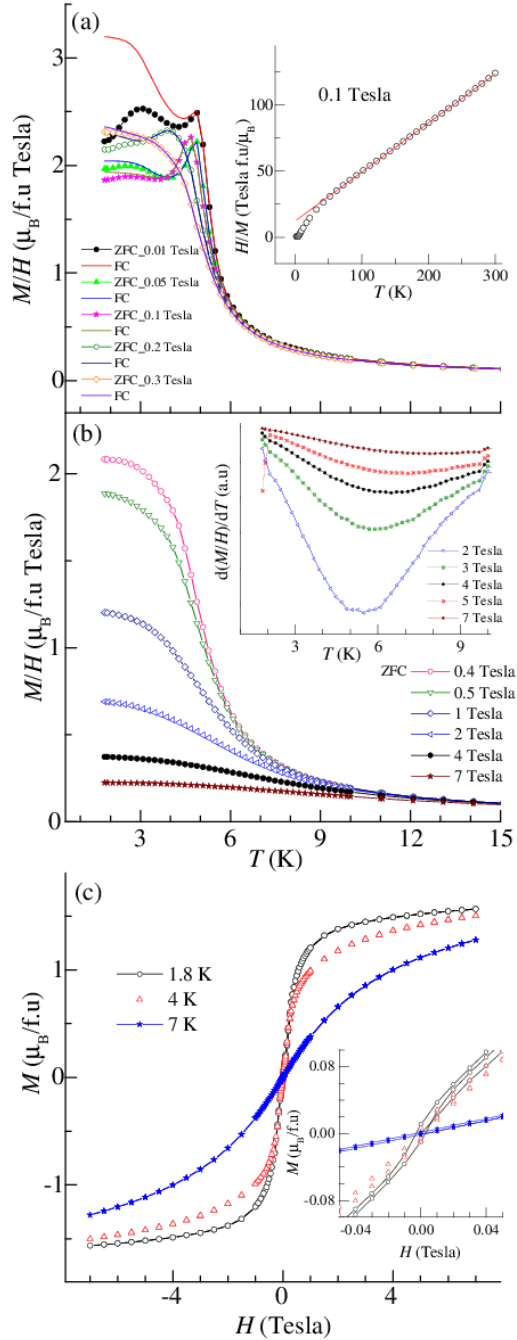

**Figure 2:** (a) Temperature dependent DC magnetization divided by magnetic field (*M*/*H*) under ZFC and FC condition in field range 0.01-0.3. Inset: Temperature dependent *H/M* at 0.1 Tesla. Red line shows the Curie Weiss law fitting. (b) Temperature dependent *M*/*H* under ZFC condition in field range 0.4-7 Tesla. Inset: Temperature dependent derivative of *M*/*H* in field range 2-7 Tesla**.** (c) Isothermal magnetization plotted as a function of magnetic field at 1.8, 4 and 7 K. Inset: Expanded form of the same curves at low fields.

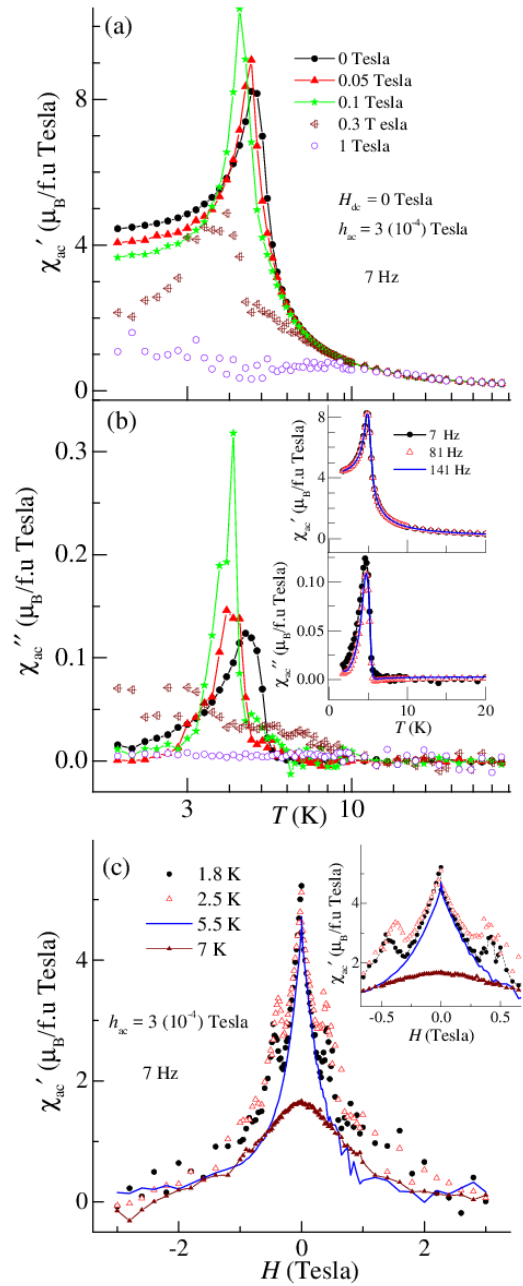

**Figure 3:** (a) Temperature dependent $\chi'_{ac}$ at different superimposed DC fields at 7 Hz in presence of 3 $(10^{-4})$ Tesla AC field. (b) Temperature dependent $\chi''_{ac}$ under similar conditions. Upper inset: Temperature dependent $\chi'_{ac}$ at different frequencies under similar conditions. Lower inset: Temperature dependent $\chi''_{ac}$ at similar frequencies under similar conditions (c) DC magnetic

field (3 to –3 Tesla) dependent $\chi'_{ac}$ at 7 Hz in presence of 3 $(10^{-4})$ Tesla AC field in temperature range 1.8-7 K. Inset: Magnified plot of the same figure at low fields.

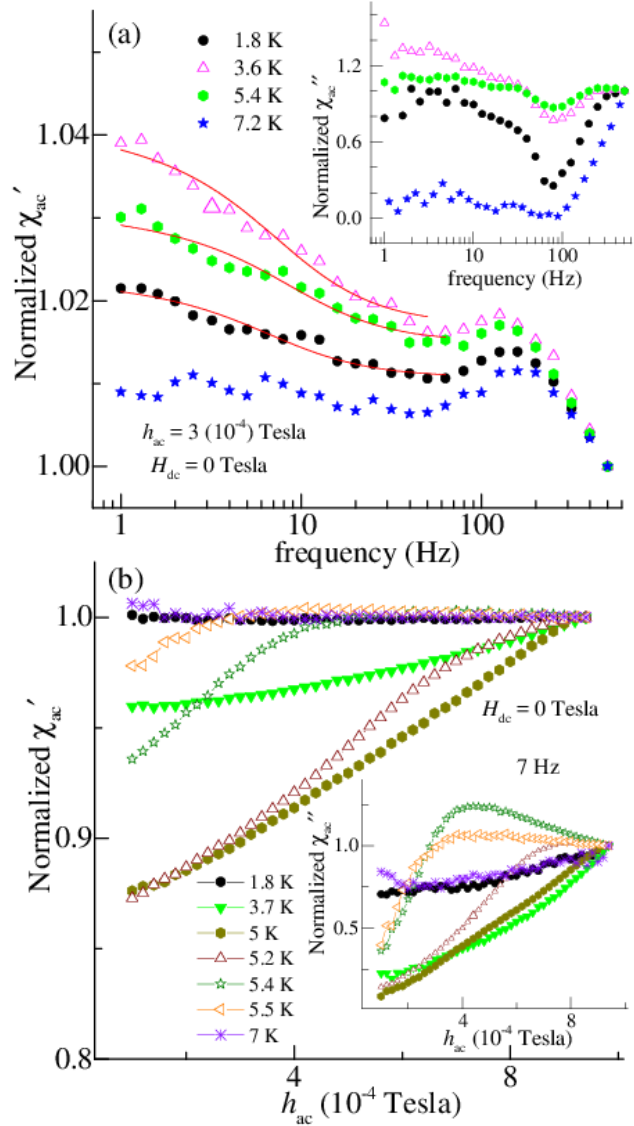

**Figure 4:** (a) Frequency dependent normalized $\chi'_{ac}$ at 0 DC field in presence of 3 $(10^{-4})$ Tesla AC field. Red curve shows the fitting of equation (2). Inset: Frequency dependent normalized $\chi''_{ac}$ under similar conditions. (b) AC field dependent normalized $\chi'_{ac}$ at 0 Tesla DC field in presence of frequency 7 Hz. Inset: AC field dependent normalized $\chi''_{ac}$ under similar condition.

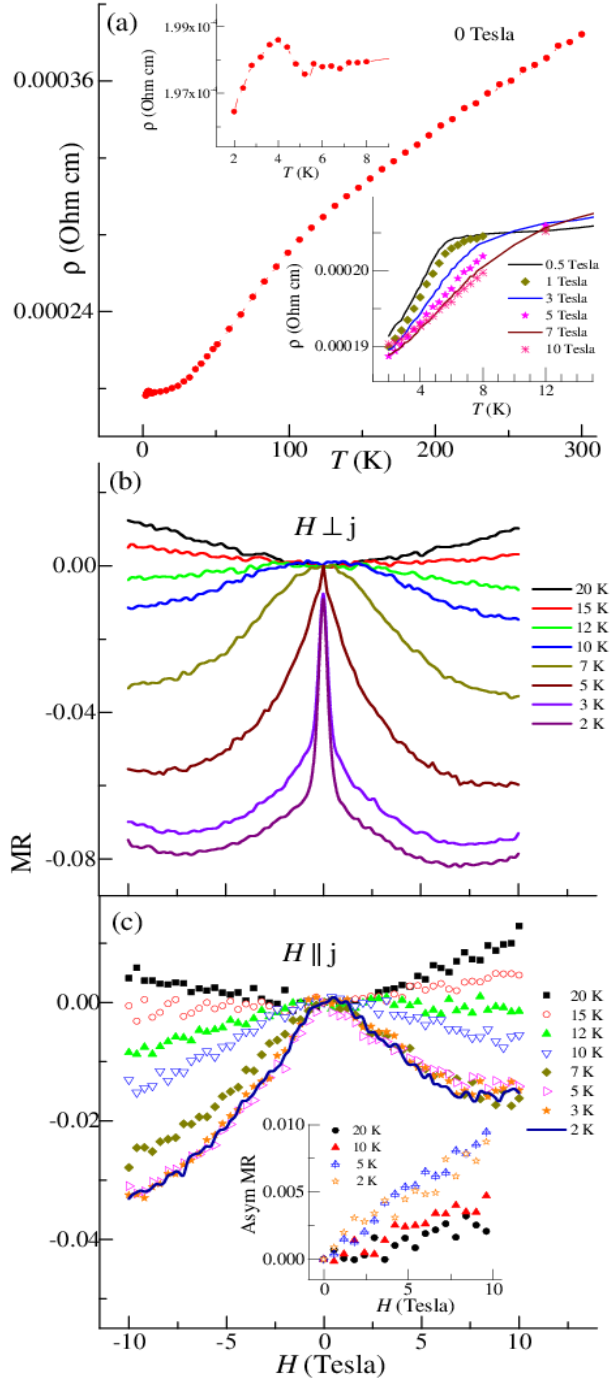

**Figure 5:** (a) Temperature dependent resistivity at 0 Tesla. Upper inset: Magnified resistivity curve in the low temperature region. Lower inset: Temperature dependent resistivity under different fields. (b) Magnetic field response of transverse MR at different temperature. (c) Magnetic field response of longitudinal MR at different temperatures. Inset: Longitudinal Asymmetric MR (Asym MR) plotted as function of magnetic field at selected temperatures.

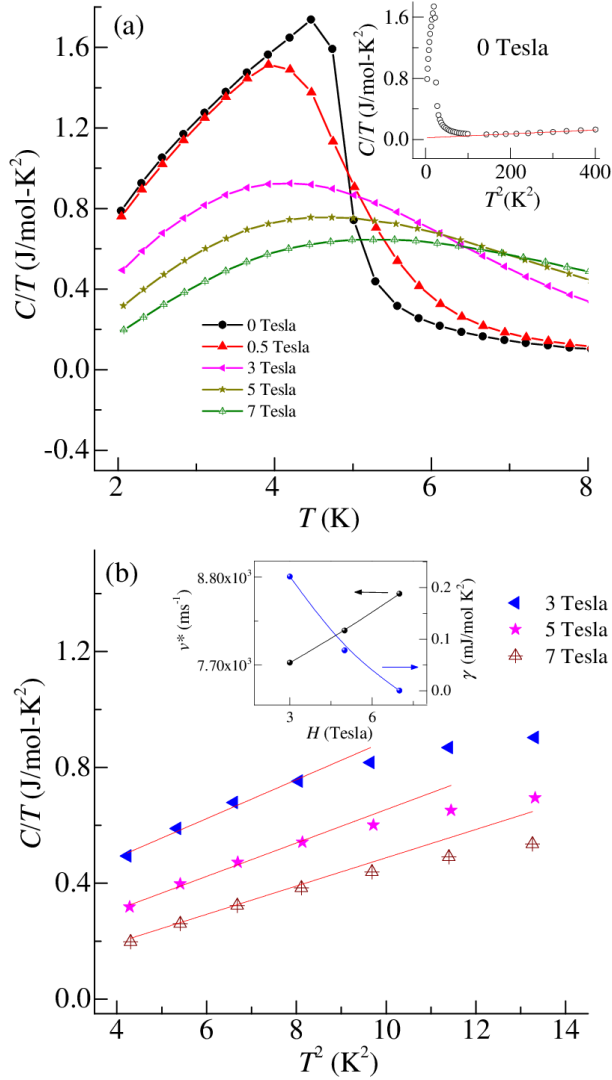

**Figure 6:** (a) Temperature response of *C*/*T* at selected magnetic fields. Upper inset: $T^2$ dependence of *C*/*T* at 0 Tesla. Solid red line is the fit of equation (7). (b) $T^2$ dependence of *C*/*T* at high fields. Solid red lines are the fit of equation (8). Inset: Effective electron velocity ($v^*$) (Left axis) and γ (Right axis) plotted as a function of magnetic field.